\documentstyle[12pt,epsf,axodraw]{article}
\begin{document}

\begin{center}
{\Large {\bf Symbolic calculation of multiparticle Feynman amplitudes}}

\vspace{0.5cm}

P.~Cherzor

{\it Skobeltsyn Institute of Nuclear Physics, Moscow State University \\
Moscow 119899, Russia }
\end{center}

\vspace{1cm}
\begin{abstract}
Using the newly modified method developed for symbolic evaluation
of Feynman amplitudes we examine two processes $2 \to 2$ (including a case
of Majorana fermions) at a tree level. Constructing special polarization
basis for spinor particles, we obtain compact expressions for helicity
amplitudes. We present the regular way for simplification of evaluated
symbolic expressions.
\end{abstract}

\vspace{1cm} 
Quantum amplitudes corresponded to Feynman diagrams depend on momenta of
particles and their polarizations. 
One can describe polarized Dirac spinors of fermion particles
in terms of helicity spinors as far as for massless fermions helicity
and chirality are connected \cite{Its-Zub}. It was developed in \cite{Kleiss85},
\cite{CALCUL81} the method to calculate polarized matrix elements
(polarized squared amplitudes) through
helicity amplitudes. A detailed description of spin formalism and
helicity spinors use one can find e.g. in \cite{Haber}. 
The amplitude is expressed as a product
of Dirac $\gamma $-matrices ended by two spinors --- {\it fermion
string}. The contribution of a fermion string to the Feynman amplitude can
be rewritten through the trace: 
\begin{equation}
\label{spinor_trace}\bar w(p_2,\lambda _2)\Gamma \;w^{\prime }(p_1,\lambda
_1)=tr(\Gamma \;w\otimes \bar w^{\prime }), 
\end{equation}
where $\Gamma $ is some product of $\gamma $-matrices.

In \cite{hep-ph/0101265} it was constructed the special spinor basis, which
allows to represent tensor products $w\otimes \bar w^{\prime }$: 
\begin{eqnarray}
\label{uu_answer}
u(p,\lambda)\otimes \bar{u}(p',\lambda') &=& N (m+\hat{p})
\left( \begin{array}{cc} 
           1-\gamma_5              & (1-\gamma_5)\hat{\eta}^* \\
          -(1+ \gamma_5)\hat{\eta} & 1+\gamma_5
       \end{array}
\right)   \hat{k} (m'+\hat{p}') \,;   \\
\label{uv_answer}
u(p,\lambda)\otimes \bar{v}(p',\lambda') &=& N (m+\hat{p})
\left( \begin{array}{cc}
         (1- \gamma_5)\hat{\eta}^* & 1-\gamma_5 \\
          1+\gamma_5               & -(1+\gamma_5)\hat{\eta}
       \end{array}
\right)  \hat{k} (m'-\hat{p}')\,;    \\
\label{vu_answer}
v(p,\lambda)\otimes \bar{u}(p',\lambda') &=& N (m-\hat{p})
\left( \begin{array}{cc}
          -(1+ \gamma_5)\hat{\eta} & 1+\gamma_5 \\ 
           1-\gamma_5              & (1-\gamma_5)\hat{\eta}^*
       \end{array}
\right) \hat{k}(m'+\hat{p}') \,;  \\
\label{vv_answer}
v(p,\lambda)\otimes \bar{v}(p',\lambda') &=& N (m-\hat{p})
\left( \begin{array}{cc} 
          1+\gamma_5               & -(1+\gamma_5)\hat{\eta} \\
         (1- \gamma_5)\hat{\eta}^* & 1-\gamma_5
       \end{array}
\right) \hat{k}(m'-\hat{p}') \,. 
\end{eqnarray}
Here $N=1/[4\sqrt{(pk)(p^{\prime }k)}]$, $m=\sqrt{p^2}$, $m^{\prime }=\sqrt{
p^{\prime 2}}$, and polarizations' dependence follows the rule: {\small $
\left( 
\begin{array}{cc}
++ & +- \\ 
-+ & -- 
\end{array}
\right) $. } The products $w\otimes \bar w^{\prime }$ with C-conjugation can
be derived by means of relations: 
\begin{equation}
\label{uvc}u(p,\pm 1)=v(p,\pm 1)^c,\,\,v(p,\pm 1)=u(p,\pm 1)^c, 
\end{equation}
\begin{equation}
\label{uv_c}\bar u(p,\pm 1)=\overline{v(p,\pm 1)^c},\,\,\bar v(p,\pm 1)=
\overline{u(p,\pm 1)^c}. 
\end{equation}
In the proposed method there were introduced three 4-vectors in Minkowsky
space: ($k$, $\eta $, $\eta ^{*}$) the same set of auxiliary vectors was
introduced for arbitrary number of fermion strings in the amplitudes. They
satisfy the relations: $(\eta ,\eta ^{*})=-1/2,\,{\eta }^2=0,\,{(\eta ^{*})}
^2=0,\,(\eta ,k)=0,\,k^2=0$.

The main features of the method we had used:

\begin{enumerate}
\item  it gives formulae uniformly applicable in massive and massless cases,
for Dirac and Majorana fermions;

\item  it can be applied to interactions with the fermion number violation;

\item  it yields more compact symbolic information on
helicity amplitudes, especially as far as the variety of polarization vectors
is reduced to three: $\eta $, $\eta ^*$, $k$.
\end{enumerate}

\noindent {\large {\bf Examples of symbolic evaluation of helicity amplitudes}}

Let us consider the processes $2\rightarrow 2$ at the tree level as an example of
evaluation helicity amplitudes. Consider $\gamma ^*$-exchange diagram for
electron-positron scattering to the quark and anti-quark $e\overline{e}
\rightarrow t\,\,\overline{t}$. 

\begin{center}
{
\unitlength=1.0 pt
\SetScale{1.0}
\SetWidth{0.7}      
\tiny    
{} \qquad\allowbreak
\begin{picture}(79,65)(0,0)
\Text(13.0,57.0)[r]{$e$}
\ArrowLine(14.0,57.0)(31.0,49.0) 
\Text(13.0,41.0)[r]{$\bar{e}$}
\ArrowLine(31.0,49.0)(14.0,41.0) 
\Text(39.0,50.0)[b]{$\gamma$}
\DashLine(31.0,49.0)(48.0,49.0){3.0} 
\Text(66.0,57.0)[l]{$t$}
\ArrowLine(48.0,49.0)(65.0,57.0) 
\Text(66.0,41.0)[l]{$\bar{t}$}
\ArrowLine(65.0,41.0)(48.0,49.0) 
\Text(39,0)[b] {diagr.1}
\end{picture} \ 
}
\end{center}

In the lowest PT order tree level 
corresponding quantum amplitude contains the fermion string: 
$$
o(\lambda _1,\lambda _2,\sigma _1,\sigma _2)=\overline{v}_{\overline{e}
}(p_1,\lambda _1)\;\gamma _\mu \;u_e(p_2,\lambda _2)\;\overline{u}
_t(q_1,\sigma _1)\;\gamma ^\mu \;v_{\overline{t}}(q_2,\sigma _2)\,\,1/s,
$$
where $p_i,\,q_j$ are particles momenta, $\lambda _i,\,\sigma _j$ are
polarizations. $1/s$ is the propagator, $s=(p_1+p_2)^2$.
One can consider electron and positron as massless, $m$ is the
mass of $t,\overline{t}$.

One can connect fermion legs in a three topologically different ways, 
each case gives the definite trace structure in the fermion string
correspondingly:

$$
o_1(\lambda _1,\lambda _2,\sigma _1,\sigma _2)=Tr\;\{\gamma _\mu
\;u_e(p_2,\lambda _2)\;\,\overline{v}_{\overline{e}}(p_1,\lambda
_1)\}\;Tr\;\{\gamma ^\mu \;v_{\overline{t}}(q_2,\sigma _2)\;\overline{u}
_t(q_1,\sigma _1)\}\,\,1/s, 
$$
$$
o_2(\lambda _1,\lambda _2,\sigma _1,\sigma _2)=Tr\;\{\gamma _\mu
\;u_e(p_2,\lambda _2)\;\overline{u}_t(q_1,\sigma _1)\;\gamma ^\mu \;v_{
\overline{t}}(q_2,\sigma _2)\;\overline{v}_{\overline{e}}(p_1,\lambda
_1)\}\,\,1/s, 
$$
$$
\,o_3(\lambda _1,\lambda _2,\sigma _1,\sigma _2)=-Tr\;\{(-\gamma _\mu )\;\,v_{
\overline{e}}^c(p_1,\lambda _1)\;\overline{u}_t(q_1,\sigma _1)\;\gamma ^\mu
\;u_{\overline{t}}^c(q_2,\sigma _2)\;\overline{u}_e^c(p_2,\lambda _2)\}\,\,1/s. 
$$
Third expression is the case of reversed spinors' order in a trace and of
presence of spinor products of C-conjugated spinors. The details of helicity
amplitudes processing algorithm, which includes a possibility to reverse the
spinors' order and the appearance of C-conjugated spinors in a trace, will
be described in \cite{algorithm}. One must replace the vertices according to 
\cite{Denner}. It was described in \cite{hep-ph/0101265} how to construct
the spinor products. For C-conjugated spinors one must use the substitutions
(\ref{uvc}), (\ref{uv_c}) in (\ref{uu_answer})-(\ref{vv_answer}). We give an
example of the result of the traces evaluation for fermion string for the
set of helicities (1,-1,-1,1), here and bellow in this section we define the
common normalization factor $N=1/16\sqrt{(p_1,k)(p_2,k)(q_1,k)(q_2,k)}$,
$N_1=N/s$: 
$$
o_1(1,-1,-1,1)=8\,N_1\,[-(k,p_1)\,(k,p_2)\,(q_1,q_2)\,-m^2\,(k,p_1)\,(k,p_2) 
$$
$$
+(k,p_1)\,(k,q_2)\,(q_1,p_2)\,-i\,(k,p_1)\,e_{\mu \nu \sigma \rho }k^\mu {p_2
}^\nu {q_1}^\sigma {q_2}^\rho  
$$
$$
+(k,p_2)\,(k,q_1)\,(p_1,q_2)\,-(k,q_1)\,(k,q_2)\,(p_1,p_2)\,-i\,(k,q_1)\,e_{
\mu \nu \sigma \rho }k^\mu {p_1}^\nu {p_2}^\sigma {q_2}^\rho ], 
$$
$$
o_2(1,-1,-1,1)=8\,N_1\,[-(k,p_1)\,(k,p_2)\,(q_1,q_2)\,-m^2\,(k,p_1)\,(k,p_2) 
$$
$$
+(k,p_1)\,(k,q_2)\,(q_1,p_2)\,-i\,(k,p_1)\,e_{\mu \nu \sigma \rho }k^\mu {p_2
}^\nu {q_1}^\sigma {q_2}^\rho  
$$
$$
+(k,p_2)\,(k,q_1)\,(p_1,q_2)\,-(k,q_1)\,(k,q_2)\,(p_1,p_2)\,-i\,(k,q_1)\,e_{
\mu \nu \sigma \rho }k^\mu {p_1}^\nu {p_2}^\sigma {q_2}^\rho ]. 
$$
Initial formula before the trace evaluation for $o_3(1,-1,-1,1)$ contains
both $\eta $ and $\eta ^{*}$ in one term.
$$
o_3(1,-1,-1,1)=4\,N_1\,[-(k,p_1)\,(k,p_2)\,(q_1,q_2)\,-2(k,p_1)\,(k,p_2)\,(\eta
,q_1)\,(\eta ^{*},q_2)\, 
$$
$$
+2(k,p_1)\,(k,p_2)\,(\eta ^{*},q_1)\,(\eta
,q_2)\,-m^2\,(k,p_1)\,(k,p_2)+(k,p_1)\,(k,q_2)\,(q_1,p_2)\, 
$$
$$
-2(k,p_1)\,(k,q_2)\,(\eta ,p_2)\,(\eta ^{*},q_1)+2(k,p_1)\,(k,q_2)\,(\eta
^{*},p_2)\,(\eta ,q_1)\,\, 
$$
$$
-2i\,(k,p_1)\,(\eta ^{*},q_1)\,e_{\mu \nu \sigma \rho }k^\mu {p_2}^\nu {q_2}
^\sigma \eta ^\rho +(k,p_2)\,(k,q_1)\,(p_1,q_2)\,+2(k,p_2)\,(k,q_1)\,(\eta
,p_1)\,(\eta ^{*},q_2)\, 
$$
$$
-2(k,p_2)\,(k,q_1)\,(\eta ^{*},p_1)\,(\eta ,q_2)+2i\,(k,p_2)\,(\eta
,q_2)\,e_{\mu \nu \sigma \rho }k^\mu {p_1}^\nu {q_1}^\sigma {(\eta ^{*})}
^\rho  
$$
$$
+2im^2\,(k,p_2)\,e_{\mu \nu \sigma \rho }k^\mu {p_1}^\nu \eta ^\sigma {(\eta
^{*})}^\rho -(k,q_1)\,(k,q_2)\,(p_1,p_2) 
$$
$$
-2(k,q_1)\,(k,q_2)\,(\eta ,p_1)\,(\eta ^{*},p_2)\,+2(k,q_1)\,(k,q_2)\,(\eta
^{*},p_1)\,(\eta ,p_2)\, 
$$
$$
+2i\,(k,q_1)\,(\eta ^{*},p_1)\,e_{\mu \nu \sigma \rho }k^\mu {p_2}^\nu {q_2}
^\sigma \eta ^\rho -2i\,(k,q_2)\,(\eta ,p_2)\,e_{\mu \nu \sigma \rho }k^\mu {
p_1}^\nu {q_1}^\sigma {\ (\eta ^{*})}^\rho ], 
$$

For all possible ways of connection of fermion legs one can select definite
expression for fermion string for each set of helicities $(\lambda
_1,\lambda _2,\sigma _1,\sigma _2)$ so, that expressions containing terms
with both $\eta $ and $\eta ^{*}$ will be always excluded \cite{algorithm}.
Doing {\it this
choice before the trace evaluation allows ones to reduce the time and
resources during trace evaluation}, because so we reduce the variety of
symbolic structures in the result of trace evaluation. Here $o_1(1,-1,-1,1)$
and $o_2(1,-1,-1,1)$ satisfy the criteria of non-appearing terms with both $
\eta $ and $\eta ^{*}$.

There exist a possibility to simplify further {\it using and processing
helicity amplitudes}. One can define, the orientation of the helicity basis
auxiliary vectors in the following way: $k^\mu =(k^{\prime },0,0,k^{\prime
}),\;\;\eta ^\mu =(0,1/2,i/2,0),\;(\eta ^{*})^\mu =(0,1/2,-i/2,0)$. The
expressions for fermion strings remains invariant to the Lorentz
transformations of initial and final particles' momenta. Only $k,\eta ,\eta
^{*}$ is fixed. One can express all terms containing $\,e_{\mu \nu \sigma
\rho }$ in the results of evaluation of fermion strings as a products of
scalar products using a component-wise representation of all vectors 
\begin{equation}
\label{conv}e_{\mu \nu \sigma \rho \;}p^\mu {q}^\nu {P}^\sigma Q^\rho =\sum
_{i,j,l,m=0,\ldots 3} ^ {i\neq j\neq l\neq m}
(p,e_i)(q,e_j)(P,e_l)(Q,e_m)\,e^{ijlm},\;\;e^{0123}=1, 
\end{equation}
$$
(p,q)=(p,e_0)(q,e_0)-(p,e_1)(q,e_1)-(p,e_2)(q,e_2)-(p,e_3)(q,e_3) 
$$
and recognizing a scalar products in this component-wise representation.

A general form of the result for fermion string is a sum over all possible
combinations of any scalar products of two external momenta and any scalar
products of one of external momenta and one of basis vectors $k,\eta ,\eta
^{*}$. Then one apply the identity (here $\{p,q\}$ are external momenta, $
\widetilde{k}$ is temporary additional auxiliary vector, satisfying $(k,
\widetilde{k})=1,\;(\eta ,\widetilde{k})=0,\;\widetilde{k}^2=0$):
$$
(p,q)=-2[(p,\eta )(q,\eta ^{*})+(p,\eta ^{*})(q,\eta )]+(p,\widetilde{k}
)(q,k)+(p,k)(q,\widetilde{k}). 
$$
After applying this identity $\widetilde{k}$ cancel themselves inside any
helicities' fermion string and do not appear in results due to the symbolic
structure of the results for helicity amplitudes we regard in this article,
because for any different external momenta $p,\,q,\,P,\,Q$ the scalar products
of any two external momenta appears only as a part of the following combination:

$$
(p,k)(q,k)(Q,P)+(P,k)(Q,k)(p,q)-(p,k)(Q,k)(q,P)-(q,k)(P,k)(p,Q). 
$$
Together with terms containing scalar products arising from convolutions 
(\ref{conv}), scalar products of external momenta expressed through scalar
products with $\eta $ and $\eta ^{*}$ contribute to the combinations of
scalar products of one of external momenta and one of basis vectors $k,\eta
,\eta ^{*}$, so that symbolic structure of external momenta $(p,q)$ can be
completely removed from the results. Non-zero results for all sets of
helicities are the following (expressions are the same for all ways of
connection of fermion legs):
$$
o'(1,-1,-1,-1)=16m(k,p_2)\,[\,(k,p_1)\,(\eta ^{*},q_1)\,+(k,p_1)\,(\eta
^{*},q_2)\,-(k,q_1)\,(\eta ^{*},p_1)\,-(k,q_2)\,(\eta ^{*},p_1)\,], 
$$
$$
o'(-1,1,-1,-1)=16m(k,p_1)\,[(k,p_2)\,(\eta ^{*},q_1)\,+(k,p_2)\,(\eta
^{*},q_2)\,-(k,q_1)\,(\eta ^{*},p_2)\,-(k,q_2)\,(\eta ^{*},p_2)\,], 
$$
$$
o'(1,-1,1,-1)=8[4(k,p_1)\,(k,p_2)\,(\eta ,q_1)\,(\eta
^{*},q_2)\,-m^2\,(k,p_1)\,(k,p_2)-4(k,p_1)\,(k,q_1)\,(\eta ,p_2)\,(\eta
^{*},q_2)\, 
$$
$$
-4(k,p_2)\,(k,q_2)\,(\eta ^{*},p_1)\,(\eta ,q_1)\,+4(k,q_1)\,(k,q_2)\,(\eta
^{*},p_1)\,(\eta ,p_2)\,], 
$$
$$
o'(-1,1,1,-1)=8[4(k,p_1)\,(k,p_2)\,(\eta ,q_1)\,(\eta
^{*},q_2)\,-m^2\,(k,p_1)\,(k,p_2) 
$$
$$
-4(k,p_1)\,(k,q_2)\,(\eta ^{*},p_2)\,(\eta ,q_1)\,-4(k,p_2)\,(k,q_1)\,(\eta
,p_1)\,(\eta ^{*},q_2)\,+4(k,q_1)\,(k,q_2)\,(\eta ,p_1)\,(\eta ^{*},p_2)\,], 
$$
$$
o'(1,-1,-1,1)=8[4(k,p_1)\,(k,p_2)\,(\eta ^{*},q_1)\,(\eta
,q_2)\,-m^2\,(k,p_1)\,(k,p_2)-4(k,p_1)\,(k,q_2)\,(\eta ,p_2)\,(\eta
^{*},q_1)\, 
$$
$$
-4(k,p_2)\,(k,q_1)\,(\eta ^{*},p_1)\,(\eta ,q_2)\,+4(k,q_1)\,(k,q_2)\,(\eta
^{*},p_1)\,(\eta ,p_2)\,], 
$$
$$
o'(-1,1,-1,1)=8[4(k,p_1)\,(k,p_2)\,(\eta ^{*},q_1)\,(\eta
,q_2)\,-m^2\,(k,p_1)\,(k,p_2) 
$$
$$
-4(k,p_1)\,(k,q_1)\,(\eta ^{*},p_2)\,(\eta ,q_2)\,-4(k,p_2)\,(k,q_2)\,(\eta
,p_1)\,(\eta ^{*},q_1)\,+4(k,q_1)\,(k,q_2)\,(\eta ,p_1)\,(\eta ^{*},p_2)\,], 
$$
$$
o'(1,-1,1,1)=16m(k,p_1)\,[-(k,p_2)\,(\eta ,q_1)\,-(k,p_2)\,(\eta
,q_2)\,+(k,q_1)\,(\eta ,p_2)\,+(k,q_2)\,(\eta ,p_2)\,], 
$$
$$
o'(-1,1,1,1)=16m(k,p_2)\,[-(k,p_1)\,(\eta ,q_1)\,-(k,p_1)\,(\eta
,q_2)\,+(k,q_1)\,(\eta ,p_1)\,+(k,q_2)\,(\eta ,p_1)\,]. 
$$
For simplicity we drop factor $N_1$:
$o'(\lambda _1,\lambda _2,\sigma _1,\sigma _2)=
o(\lambda _1,\lambda _2,\sigma _1,\sigma _2)/N_1$
in the above expressions.
Here we observe that scalar products of two external momenta do not appear
in the result.

One may represent the results in a shorter form. Collect all scalar products
to $\,e_{\mu \nu \sigma \rho }$-structures using the identities: 
\begin{equation}
\label{iden}\Delta (p)=i/2\,(p,k)=\,e_{\mu \nu \sigma \rho }k^\mu \,p^\nu
\eta ^\sigma (\eta ^{*})^\rho , 
\end{equation}
\begin{equation}
\label{16'}\Lambda (p,q)=i[(\eta ,p)\,(q,k)-(\eta ,q)\,(p,k)]=\,\,e_{\mu \nu
\sigma \rho }k^\mu \,p^\nu q^\sigma \eta ^\rho , 
\end{equation}
\begin{equation}
\label{16}{\Lambda ^{*}}(p,q)=i[(p,k)\,(\eta ^{*},q)-(p,\eta
^{*})\,(q,k)]=\,e_{\mu \nu \sigma \rho }k^\mu \,p^\nu q^\sigma (\eta
^{*})^\rho , 
\end{equation}
Non-zero results for helicity amplitudes became the following:%
$$
o(1,-1,-1,-1)=N_1\,32\,m\,\Delta (p_2)\,\,\,{\Lambda ^{*}}((q_1+q_2),p_1)\,, 
$$
$$
o(-1,1,-1,-1)=N_1\,32\,m\,\Delta (p_1)\;{\Lambda ^{*}}((q_1+q_2),p_2), 
$$
$$
o(1,-1,1,-1)=N_1\,32\,[\,{\Lambda ^{*}}(p_1,q_2)\,\,\Lambda (p_2,q_1)+\,m^2\Delta
(p_1)\,\Delta (p_2)], 
$$
$$
o(-1,1,1,-1)=N_1\,32\,[\,{\Lambda ^{*}}(p_2,q_2)\,\,\Lambda (p_1,q_1)+\,m^2\Delta
(p_1)\,\Delta (p_2)], 
$$
$$
o(1,-1,-1,1)=N_1\,32\,[{\Lambda ^{*}}(p_1,q_1)\;\Lambda (p_2,q_2)+\,m^2\Delta
(p_1)\,\Delta (p_2)], 
$$
$$
o(-1,1,-1,1)=N_1\,32\,[{\Lambda ^{*}}(p_2,q_1)\,\,\Lambda (p_1,q_2)+\,m^2\Delta
(p_1)\,\Delta (p_2)], 
$$
$$
o(1,-1,1,1)=N_1\,32\,m\,\Delta (p_1)\,\Lambda ((q_1+q_2),p_2)\,, 
$$
$$
o(-1,1,1,1)=N_1\,32\,m\,\Delta (p_2)\,\Lambda ((q_1+q_2),p_1). 
$$
Our result for helicity amplitudes for selected diagram do not
contradict to the results obtained earlier by other researchers
(compare e.g. \cite{4}). Although we did use helicity basis that differs from
one of \cite{4}, so the straightforward symbolic comparison fails.
Indeed the value of the helicity amplitude depends on the method with which
it was built. Amplitude in general case could have a complex part.
The values those can be compared are polarized matrix elements (squared amplitudes).
On the initial stage of our research we did not apply detailed comparison methods
to check agreement with other researches done. Instead we concentrated
ourself, first, on check the correctness of the results comparing to non-polarized
matrix element that evaluates in a well-known way and the correctness
under transformations of helicity basis, spinor substitutions due
to C-conjugation symmetry, changes trace structure applied.
To make a numerical tests we assigned the numerical values (allowed by
conservation and Minkowsky space symmetry laws) to particles' momenta
and masses.

As an example with Majorana fermions we had examined the process $e\overline{e
}\rightarrow \,neutralino\;pair$ 
in the lowest order tree level with the same procedure of traces
optimization, here we also considered electron and positron as massless.
This is a process that includes particles neutralino and anti-neutralino,
they transfer one into the other by C-conjugation operation.

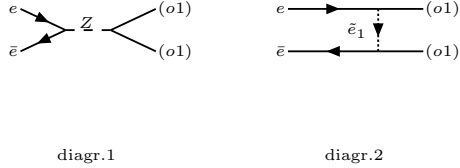
\begin{center}
{
\unitlength=1.0 pt
\SetScale{1.0}
\SetWidth{0.7}      
\tiny    
{} \qquad\allowbreak
\begin{picture}(79,65)(0,0)
\Text(13.0,57.0)[r]{$e$}
\ArrowLine(14.0,57.0)(31.0,49.0) 
\Text(13.0,41.0)[r]{$\bar{e}$}
\ArrowLine(31.0,49.0)(14.0,41.0) 
\Text(39.0,50.0)[b]{$Z$}
\DashLine(31.0,49.0)(48.0,49.0){3.0} 
\Text(66.0,57.0)[l]{$(o1)$}
\Line(48.0,49.0)(65.0,57.0) 
\Text(66.0,41.0)[l]{$(o1)$}
\Line(48.0,49.0)(65.0,41.0) 
\Text(39,0)[b] {diagr.1}
\end{picture} \ 
{} \qquad\allowbreak
\begin{picture}(79,65)(0,0)
\Text(13.0,57.0)[r]{$e$}
\ArrowLine(14.0,57.0)(48.0,57.0) 
\Text(66.0,57.0)[l]{$(o1)$}
\Line(48.0,57.0)(65.0,57.0) 
\Text(44.0,49.0)[r]{$\tilde{e}_1$}
\DashArrowLine(48.0,57.0)(48.0,41.0){1.0} 
\Text(13.0,41.0)[r]{$\bar{e}$}
\ArrowLine(48.0,41.0)(14.0,41.0) 
\Text(66.0,41.0)[l]{$(o1)$}
\Line(48.0,41.0)(65.0,41.0) 
\Text(39,0)[b] {diagr.2}
\end{picture} \ 
}
\end{center}

One can write amplitude for the diagram with Z-boson at s-channel in a following form:
$$
f(\lambda _1,\lambda _2,\sigma _1,\sigma _2)=\overline {v}_{\overline e}(p_1,
\lambda _1)\,\Gamma _{\mu }\,u_e(p_2,\lambda _2)\,
\overline{u}_{\widetilde {\chi }_1^0}(q_1,\sigma _1)\,
\Gamma ^{\prime \mu }\,v_{\widetilde {\chi }_1^{0\,(c)}}(q_2,\sigma _2)\,\,1/(s-{m_Z}^2),
$$
where
$$
\Gamma _\mu =\cos (2\theta _w)\,\gamma _\mu \,(1-\gamma ^5)-2(\sin \theta
_w)^2\,\gamma _\mu \,(1+\gamma ^5),\;\Gamma _\mu ^{\prime }=c_1\,\gamma _\mu
\,\gamma ^5 
$$
is the vertices structure. $m_Z$ is the $Z$-bozon mass.
For another diagram with s-electron at t-channel:
$$
v(\lambda _1,\lambda _2,\sigma _1,\sigma _2)=\overline {v}_{\overline e}(p_1,
\lambda _1)\,\Pi \,v_{\widetilde {\chi }_1^{0\,(c)}}(q_2,\sigma _2)\,
\overline{u}_{\widetilde {\chi }_1^0}(q_1,\sigma _1)\,{\Pi }'\,
u_e(p_2,\lambda _2)\,\,1/(t-{m_{\tilde {e}}}^2).
$$
$\Pi =c_2\,(1+\gamma ^5)$, ${\Pi }'=c_2\,(1-\gamma ^5)$ are vertices. 
$c_1$, $c_2$ depend on SUSY parameters, $t$ -- Mandelstam variable, $m_{\tilde {e}}$ --
selectron mass.

In this case one have a possibility to make a fermion lines conjunction of
Feynman diagrams in three topologically inequivalent ways for both diagrams.
We apply formulae:
$$
f_1(\lambda _1,\lambda _2,\sigma _1,\sigma _2)=Tr\;\{\Gamma _\mu \,
u_e(p_2,\lambda _2)\,\overline{u}_{\widetilde{\chi }
_1^0}(q_1,\sigma _1)\,\Gamma ^{^{\prime }\mu }\,v_{\widetilde{\chi }
_1^{0\,(c)}}(q_2,\sigma _2)\,
\overline{v}_{\overline{e}}(p_1,\lambda_1)\}\,\,1/(s-{m_Z}^2), 
$$
$$
f_2(\lambda _1,\lambda _2,\sigma _1,\sigma _2)=Tr\;\{\Gamma _\mu \,
u_e(p_2,\lambda _2)\,\overline{v}_{\overline{e}}(p_1,\lambda _1)
\}\;Tr\;\{\Gamma ^{^{\prime }\mu }\,v_{
\widetilde{\chi }_1^{0\,(c)}}(q_2,\sigma _2)\,\overline{u}_{\widetilde{\chi }
_1^0}(q_1,\sigma _1)\}\,\,1/(s-{m_Z}^2), 
$$
$$
f_3(\lambda _1,\lambda _2,\sigma _1,\sigma _2)=-Tr\;\{\widetilde{
\Gamma }_\mu \,v_{\overline{e}}^c(p_1,\lambda _1)\,\overline{u}_{\widetilde{
\chi }_1^0}(q_1,\sigma _1)\,\Gamma ^{^{\prime }\mu }\,v_{\widetilde{\chi }
_1^{0\,(c)}}(q_2,\sigma _2)\,\overline{u}_e^c(p_2,\lambda _2)\}\,\,1/(s-{m_Z}^2), 
$$
where one obtains $\widetilde{\Gamma }_\mu$ form ${\Gamma }_\mu$ using rules 
\cite {Denner}. $N_2=N/(s-{m_Z}^2)$.
$$
v_1(\lambda _1,\lambda _2,\sigma _1,\sigma _2)=Tr\;\{\Pi \;v_{\widetilde{\chi 
}_1^{0\,(c)}}(q_2,\sigma _2)\;\overline{u}_{\widetilde{\chi }%
_1^0}(q_1,\sigma _1)\;\Pi ^{\prime }\,u_e(p_2,\lambda _2)\,\,\overline{v}_{
\overline{e}}(p_1,\lambda _1)\}\,\,1/(t-{m_{\tilde {e}}}^2), 
$$
$$
v_2(\lambda _1,\lambda _2,\sigma _1,\sigma _2)=Tr\;\{\Pi \;v_{\widetilde{\chi 
}_1^{0\,(c)}}(q_2,\sigma _2)\,\overline{v}_{\overline{e}}(p_1,\lambda
_1)\}\;Tr\;\{\Pi ^{\prime }\;u_e(p_2,\lambda _2)\;\overline{u}_{\widetilde{
\chi }_1^0}(q_1,\sigma _1)\}\,\,1/(t-{m_{\tilde {e}}}^2), 
$$
$$
v_3(\lambda _1,\lambda _2,\sigma _1,\sigma _2)=-Tr\;\{\Pi \;v_{\overline{e}
}^c(p_1,\lambda _1)\;\overline{u}_{\widetilde{\chi }_1^0}(q_1,\sigma
_1)\;\Pi ^{\prime }\;u_e(p_2,\lambda _2)\;\overline{v}_{\widetilde{\chi }
_1^{0\,(c)}}^c(q_2,\sigma _2)\}\,\,1/(t-{m_{\tilde {e}}}^2). 
$$
Reversing a direction of fermion line in $v_3(\lambda _1,\lambda _2,
\sigma _1,\sigma _2)$ does not change a vertices structure because vertices do
not contain ${\gamma }_{\mu }$ but ${\gamma }_5$.
In cases of C-conjugated spinors we had implied the
formulae (\ref{uu_answer})-(\ref{vv_answer}) with substitutions (\ref{uvc}),
(\ref{uv_c}). We had find the same symbolic structures after trace
evaluation as in non-Majorana case, and had applied the same procedure of
optimization symbolic results. We show here results (non-zero
expressions) one obtains after simplification for helicity amplitudes fermion strings
for the diagram with Z-boson at s-channel:
$$
f(1,-1,-1,-1)=N_2\,64\,c_1\,\cos (2\theta _w)\,m\,\Delta (p_2)\,
\Lambda ^*(q_2-q_1,p_1),
$$
$$
f(-1,1,-1,-1)=-N_2\,64\,(2\,\sin ^2\theta _w)\,m\,\Delta (p_1)\,
\Lambda ^*(q_2-q_1,p_2),
$$
$$
f(1,-1,1,-1)=N_2\,64\,c_1\,\cos(2\theta _w)\,[\Lambda ^*(p_1,q_2)\,
\Lambda (p_2,q_1)-m^2\,\Delta (p_1)\,\Delta (p_2)],
$$
$$
f(-1,1,1,-1)=-N_2\,64\,c_1\,(2\,\sin ^2\theta _w)\,
[\Lambda ^*(p_2,q_2)\,\Lambda (p_1,q_1)-m^2\,\Delta (p_1)\,\Delta (p_2)],
$$
$$
f(1,-1,-1,1)=N_2\,64\,c_1\,\cos(2\theta _w)\,[-\Lambda ^*(p_1,q_1)\,
\Lambda (p_2,q_2)+m^2\,\Delta (p_1)\,\Delta (p_2)],
$$
$$
f(-1,1,-1,1)=-N_2\,64\,c_1\,(2\,\sin ^2\theta _w)\,
[-\Lambda ^*(p_2,q_1)\,\Lambda (p_1,q_2)+m^2\,\Delta (p_1)\,\Delta (p_2)],
$$
$$
f(1,-1,1,1)=N_2\,64\,c_1\,\cos (2\theta _w)\,m\,\Delta (p_1)\,\Lambda (q_2-q_1,p_2),
$$
$$
f(-1,1,1,1)=-N_2\,64\,c_1\,(2\,\sin ^2\theta _w)\,m\,\Delta (p_2)\,
\Lambda (q_2-q_1,p_1),
$$
and for the diagram with selectron at t-channel:
$$
v(1,-1,-1,-1)=N_3\,64\,{c_2}^2\,m\,\Delta (p_2)\,{\Lambda ^{*}}
(q_2,p_1),\;\;v(1,-1,1,-1)=N_3\,64\,{c_2}^2\,{\Lambda ^{*}}(p_1,q_2)\,{\Lambda }
(p_2,q_1), 
$$
$$
v(1,-1,-1,1)=N_3\,64\,{c_2}^2\,m^2\,\Delta (p_1)\,\Delta
(p_2),\;\;\;v(1,-1,1,1)=N_3\,64\,{c_2}^2m\,\Delta (p_1)\,{\Lambda }(q_1,p_2). 
$$
Variable $m$ is the neutralinos' mass, $N_3=N/(t-{m_{\tilde {e}}}^2)$.
One can see that chiral symmetry in vertices factors gives the influence
on the structure of helicity amplitudes.\\

\noindent {\large {\bf Conclusions}}

We have considered two examples of
evaluation helicity amplitudes for the tree level for the processes $
2\rightarrow 2$ (including the case of Majorana external particles).

The key point of our approach is the fixation of three polarization basis
vectors --- helicity basis --- the same for each spinor pair.
In case of use massive spinors if to be precise one should keep the orientation
of the polarization vectors that is in general not coincide with the directions
of ones for massless spinors. In our consideration we left the directions of
polarization vectors as for massless case indicating the mass presence
outside of polarization vectors part. See (\ref{uu_answer})-(\ref{vv_answer}).
The above approach leads to significant simplifications in case of multi-particle
production.
The recipe of traces structure optimization before traces evaluation proposed
in \cite{hep-ph/0101265} was implied successfully (see \cite{algorithm}).

After fixation the helicity basis the further optimization of symbolic
structure of the results is achieved for regarded scattering processes. At the first
step of optimization the results had been written in a form of scalar
products. Symbolic results for pre-optimized helicity amplitudes obtained in
this form do not depend on the order of connection of external particles
lines. This feature serves for verification the correctness of symbolic expressions
for helicity amplitudes.

Scalar products in the considered helicity fermion strings could be
completely factorized in the form of special $\Delta ,\,\Lambda ,{\Lambda
^{*}}$ symbolic structures. The last could be also a regular feature,
possibly it remains in the scattering $2\rightarrow n$ at least in some
cases of the traces' structure. A reason to think so is: at least in the
case of real particles one may treat the fermion string for the diagram as a
number of traces: $Tr\;\{\Gamma \,w_i(p_1,\lambda _1)\;\overline{w}
_j(p_2,\lambda _2)\}$, where $\Gamma $ is a vertex (or a construction that
consists of some vertices and lines either internal (propagators of virtual
particles) or external (boson lines
e.g.)) 
and $w$ are the spinors, and each such trace has a form, we can
observe in the case $2\rightarrow 2$ in a sense of spinors structure, but
all non-trivial symbolic structures (containing spinor particles'
polarization vectors) arises exactly from spinor products. The difficulties
may arise when we will deal with: virtual particles, exotic expressions for
vertices, external boson lines, so that the $\Gamma $ structure will be complicate.

The procedure implied for symbolic optimization of regarded processes'
helicity amplitudes could be extended to the case of more complicated processes.
Possibly it may occur rather economical to operate with results for helicity
amplitudes in symbolic form, then to store all terms for usual non-polarized
cross-sections for the calculations in the case of scattering $2\rightarrow
n $, where $n>4$.

As the next step it could be considered how to apply more advanced tests
for our helicity amplitudes' results and comparison methods with the results obtained by
the other reseachers in application for calculation polarized matrix elements
for the elementary particles' scattering processes.\\

\noindent {\large {\bf Acknowledgments}}

This work is supported by grants: RFFI 01-02-16710, INTAS CERN99-0377.

Sincere thanks to prof. A.~Pukhov for fruitful discussions and important
support and to prof. V.~Ilyin for discussions and valuable support during work on project
at Skobeltsyn Institute of Nuclear Physics, Moscow State University.
\nocite{*} \bibliographystyle{aipproc}

\end{document}